# Profile Monitor SEM's for the NuMI Beam at Fermilab

Dharmaraj Indurthy, Sacha Kopp, Marek Proga, Zarko Pavlovich

*Department of Physics, University of Texas at Austin,*
*Austin, Texas 78712*

**Abstract.** The Neutrinos at the Main Injector (NuMI) project will extract 120 GeV protons from the FNAL Main Injector in 8.56 μsec spills of $4 \times 10^{13}$ protons every 1.9 sec. We have designed secondary emission monitor (SEM) detectors to measure beam profile and halo along the proton beam transport line. The SEM's are Ti foils 5μm in thickness segmented in either 1 mm or 0.5 mm pitch strips, resulting in beam loss $\sim 5 \times 10^{-6}$. We discuss aspects of the mechanical design, calculations of expected beam heating, and results of a beam test at the 8 GeV transport line to MiniBoone at FNAL.

## INTRODUCTION

The NuMI beamline [1,2] at Fermilab will deliver an intense muon neutrino beam to the MINOS detectors at FNAL and at the Soudan Laboratory in Minnesota. The primary beam is fast-extracted onto the NuMI pion production target in 8.56μsec spills from the 120 GeV FNAL Main Injector. The beam line is designed to accept $4 \times 10^{13}$ protons/spill. Along most of the transport line, the spot size is 1-2 mm. Because of the large 400kW average beam power, beam loss must be kept to of $<10^{-5}$ in order to keep residual activation of beam components to a manageable level and in order to prevent activation of water from aquifers near the underground NuMI tunnel.

Instrumentation along the 700m primary beam line is to include beam profile monitors, capacitive beam position monitors (BPM's), a resistive wall monitor, and two beam current toroids (BCT's). Additionally, the NuMI target is electrically isolated, so may be read out as a Budal-type monitor [3], and an ion chamber array near the beam dump measures the remnant primary beam after a long 725m free drift, so acts as an accurate measure of the final proton beam angle [4]. The intercepting profile monitor SEM's serve as the initial calibration of the BPM's and ion chamber.

Some of the requirements of the profile monitor SEM's are as follows: (1) the device must survive $\sim 10^{20}$ particles/cm$^2$ per year; (2) groundwater activation in the carrier tunnel region requires beam loss $\sim 10^{-6}$; (3) the beam centroid position accuracy must be ~100μm along the transport line and ~50μm for the pre-target SEM's; (4) the SEM chamber vacuum must be $10^{-8}$ Torr; (5) the measurement aperture must be ~2" and clear aperture 4"; (6) the device must be removable from the beam without turning off the beam, and replacing the device in the beam must be achieved to 50μm positional accuracy.

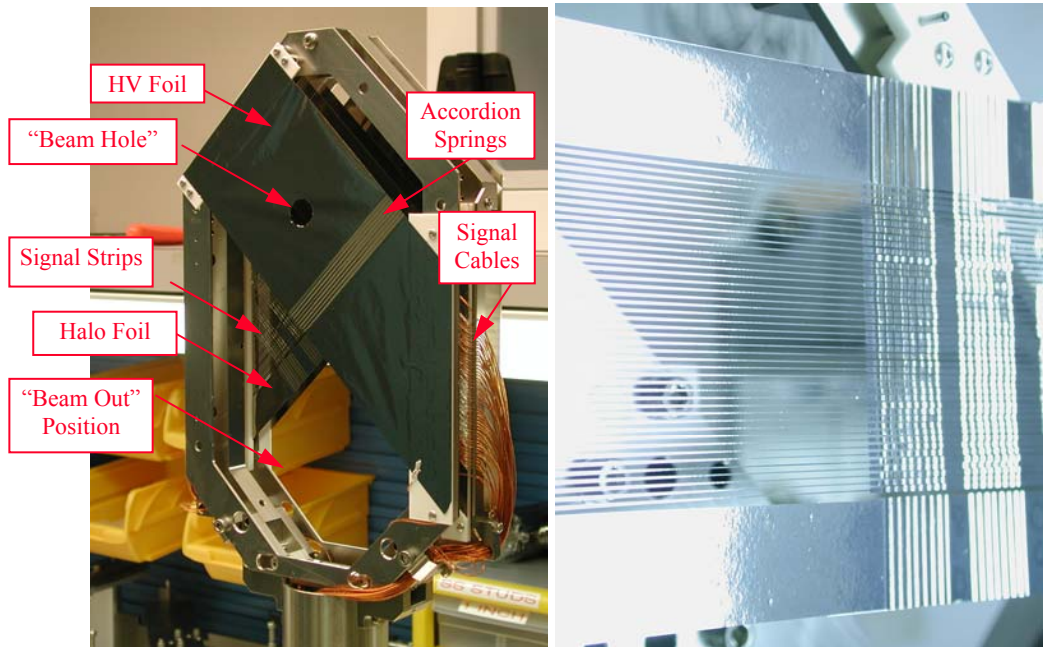

**FIGURE 1.** (left) Photograph of the foil paddle. Five foils are mounted, 3 bias foils and two segmented foils for *X* and *Y* profiles of the beam. The paddle surrounds the beam at all times, so the foils may be inserted or retracted from the beam without the paddle passing through the beam. (right) Close-up view of one segmented Ti foil. The segmented strips are at 1 mm pitch and measure out to 2.2cm radius. A wide strip measures beam halo out to 3.8cm radius. Accordion springs pressed into the signal and HV foils maintain tension against beam heating. The strips are 0.75 mm wide at the ends, where accordion springs are pressed in to maintain strip tension. In the area traversed by the beam, the strips are 0.15 mm wide.

## PROFILE MONITOR SEM DESIGN

We have borrowed from a foil SEM at CERN [5] and designed a large-aperture SEM constructed from 5μm Ti foils. Titanium is selected because it appears to maintain its secondary electron yield even after large particle fluences [6]. This is in contrast to the standard "multiwire" SEM's in use at FNAL which utilize Au-plated W-Rh wires, which were observed to degrade in signal yield by more than 20% over the course of fixed target operations to KTeV [7].

A view of the foil SEM paddle is shown in Figure 1. Three planes of solid Ti foil, 2.5μm in thickness, are interleaved with segmented foils intended for *X* and *Y* profiles. The signal foils are 5μm thick. In this figure, the strip pitch is 1 mm. Over the 8cm diameter area of the SEM traversed by the beam, the signal foil strips are 0.15 mm in width. A total of 44 strips are etched into the foil, along with 1.5cm wide strips to to measure beam halo out to 3.8cm radius. The paddle completely surrounds the beam, so the foils may be inserted into the beam or retracted without the paddle frame traversing the beam. The foils are mounted on precise ceramic combs which define the strip pitch and the foil location on the paddle [8]. A "beam hole" of 12 mm diameter in the bias foil permits most of the 1 mm beam spot to pass through without beam loss, while still permitting adequate voltage bias to maintain signal yield.

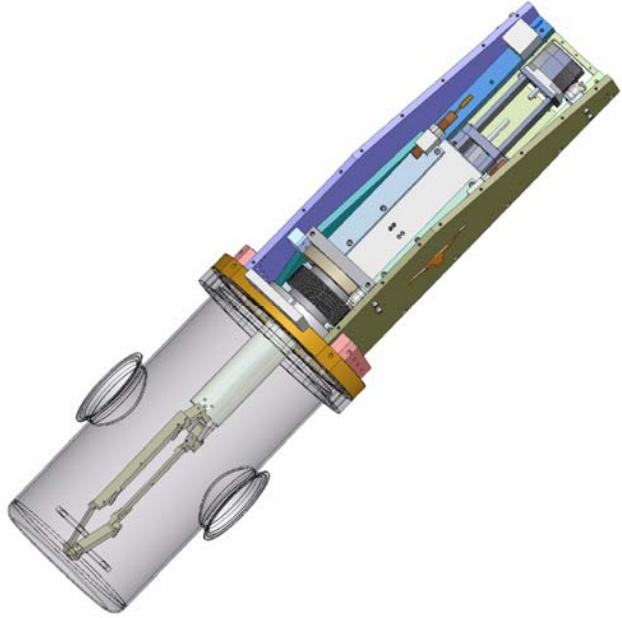

**FIGURE 2.** Schematic drawing of the foil SEM vacuum can and actuator assembly. The beam ports are 10cm diameter quick-disconnect flanges. The actuator and foil paddle are mounted on a 25cm diameter flange. The actuator consists of a stepper motor and linear stage driving a vacuum bellows to which is rigidly connected a 5cm hollow shaft. The foil paddle is cantilevered off the shaft.

Each signal strip has 32 accordion springs pressed into its ends. The springs are elongated by approximately 4 mm prior to installation on the paddle. At this extension, our measurements indicate a tension of ~1g is attained on each strip. The mechanical sag $\delta y$ of a strip of cross-sectional area $A$, length $L$, density $\rho$, may be approximated as $\delta y = g\rho A L^2/T$. This expression gives, for $L$=12 cm, a tension $T$=1.0 g, the density of Titanium $\rho$=4.5 g/cm$^3$, and a 5 μm × 0.15 mm strip, a sag $\delta y$=40 μm. Digital photographs of the mounted signal foils confirm a pitch of 1.0 mm, with strip-to-strip variation <15μm (the resolution of the photographs is 15 μm).

The foil actuating mechanism and vacuum can are shown in Figure 2. The foil paddle is cantilevered on a 5cm diameter hollow shaft, the other end of which is welded to a 12cm "conflat" flange. A linear motion stage and stepper motor actuate the assembly into or out of the beam. A 9cm outer diameter, 6.4cm inner diameter bellows forms the vacuum seal for the actuator. Ceramic-insulated limit switches [9] halt the stepper travel at either end, while a 1.25 mm range linear variable differential transformer (LVDT) confirms the final beam "in" position of the foils with 1μm accuracy [10]. Kapton-insulated signal cables are routed through the hollow shaft to feedthroughs at its end. The cables are bonded to the foil strips using a conducting epoxy appropriate for ultra-high vacuum [11], which is also how the strips are bonded to the ceramics. Brackets mount the linear stage to the large 25cm diameter conflate flange on the end of the SEM's vacuum chamber. Precise dowel holes in the 25cm flange and in the moving flange at the end of the bellows allow *in situ* optical survey of the foil position when installed in the beam line. The 16 liter vacuum chamber is 20cm diameter cylinder with 10cm diameter quick-disconnect flanges at the beam ports.

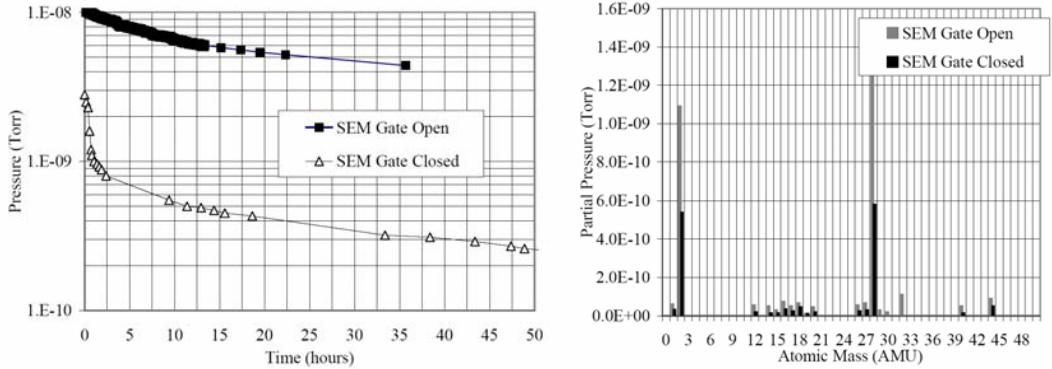

**FIGURE 3.** Vacuum tests of completed a SEM. Both measurements were performed on a vacuum system pumped on by a 30 liter/sec ion pump, with a gate valve which could isolate or include the SEM chamber. (left) Pump-down curves. (right) Residual gas analysis.

Tests have been performed of the linear motion actuator's repeatability. These tests include tests under vacuum load and in air. We find the repeatability is better than 15 µm over 24-48 hour periods, consistent with temperature variations in the laboratory and the differential thermal expansion of the materials in the actuator.

Vacuum tests were performed on a system which could isolate or include the SEM chamber via a UHV gate valve between the SEM and the rest of the system. When the SEM was isolated, a rate-of-rise measurement indicates an outgas rate of $1.2 \times 10^{-7}$ Torr-liter/sec. This outgas rate, furthermore, is consistent with the ultimate pressure achieved with the SEM included in the vacuum system (see Figure 3).

## THERMAL SIMULATIONS

We have performed detailed finite element calculations of the temperature induced in the foil SEM's due to heating by the NuMI beam [12]. The heat input to the SEM comes from the energy lost by the beam, and the power dissipated by the SEM comes from blackbody radiation and from thermal conduction through the SEM material. We performed the calculations for several materials using thermal conductivity, heat capacity, and density values from Ref. [13]. We also compared the heating of wires and foils; wire SEM's cool less efficiently because blackbody radiation is proportional to the surface area of the emitter.

The temperature rise in the SEM material results from ionization energy loss by the 120 GeV protons. We use the "restricted energy loss", which accounts for ionization leading to electrons less than a cutoff energy $T_{cut}$: $dE/dx(T_e < T_{cut})$ [14]. For the cutoff energy $T_{cut}$, we use the energy at which the range of an electron is ½ the thickness of the SEM material. The range values are taken from Ref [15]. This restricted $dE/dx$ accounts properly for the fact that some δ rays escape out the back of the SEM, so do not deposit their energy in the SEM. The restricted $dE/dx$ values are given in Table 1. The effect of restricted energy loss is greater for thin foils than for wire SEM's and tends to lower the predicted energy deposited in the SEM, so this effect tends to again yield larger temperature predictions for a wire SEM as compared to a foil SEM.

**TABLE 1.** Results of thermal simulations for foil and wire SEM's.

| Element | 5μm thick foil | | | | 50μm diameter wire | | | | Yield Stress (MPa) |
|---|---|---|---|---|---|---|---|---|---|
| | dE/dx (MeV cm²/g) | ΔT (°C) | ΔL (μm) | S (MPa) | dE/dx (MeV cm²/g) | ΔT (°C) | ΔL (μm) | S (MPa) | |
| Be | 1.14 | 64 | 11 | 188 | 1.23 | 69 | 20 | 202 | 240 |
| C | 1.26 | 186 | 1.3 | 19 | 1.37 | 202 | 3 | 20 | 469 |
| Al | 1.22 | 138 | 35 | 238 | 1.32 | 150 | 66 | 258 | 10-35 |
| Ti | 1.14 | 222 | 12 | 208 | 1.23 | 240 | 41 | 224 | 220 |
| Ni | 1.15 | 264 | 54 | 736 | 1.25 | 287 | 118 | 800 | 1580 |
| Ag | 1.05 | 449 | 66 | 709 | 1.13 | 484 | 106 | 763 | |
| W | 0.94 | 715 | 30 | 1305 | 1.02 | 775 | 64 | 1416 | 550 |
| Au | 0.93 | 729 | 84 | 807 | 1.01 | 791 | 154 | 877 | 205 |

Figure 4 shows the results of the thermal model for a Ti foil 5 μm in thickness. The left plot of Figure 4 shows the temperature profile along the center-most strip at several times during the 1.9sec beam cycle, after many transpired beam cycles. As seen in the graph, the beam causes a sharp rise in temperature at $t=0$ sec. The cooling between spills is predominantly due to blackbody radiation. The (small) effect of thermal conduction is evident by the broadening temperature profile over the course of the cooling cycle. The right graph of Figure 4 shows the linear expansion of the foil strip, and compares to the elongation expected for a Ti wire 50 μm in diameter.

Table 1 provides the temperature rise $\Delta T$, maximum linear elongation $\Delta L$, as well as dynamic stress $S=\alpha E\Delta T$, where $\alpha$ is the coefficient of linear expansion and $E$ the Young's Modulus. The dynamic stress may be compared to the "yield stress" for the material, the point at which the material may deform plastically. Beryllium, Carbon fiber and Titanium are preferable from the point of view of long term material damage, *ie*: have dynamic stress values below the yield stress.

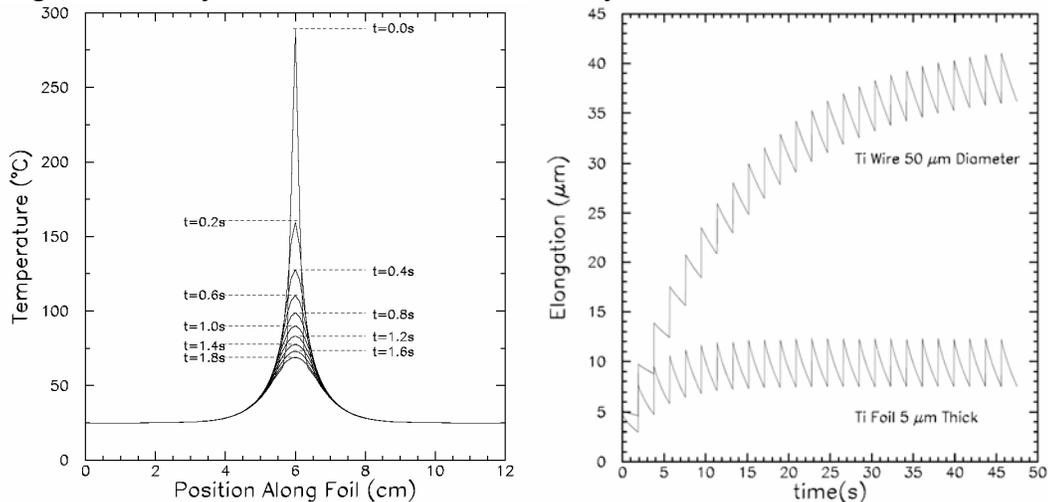

**FIGURE 4.** Results of the thermal model of 5μm Ti foil in the NuMI beam. (left) Temperature along the center most strip at several time increments through one beam cycle: at $t=0$sec the beam passes through the foil, and at $t=1.8$ sec the foil has cooled to its minimum temperature, just prior to the next spill. (right) Net elongation of a 12cm long, 5μm thick, Ti foil strip as a function of time, showing the repeated heating and cooling of several beam cycles. The Ti foil's elongation is compared to that for a 50 μm diameter Ti wire.

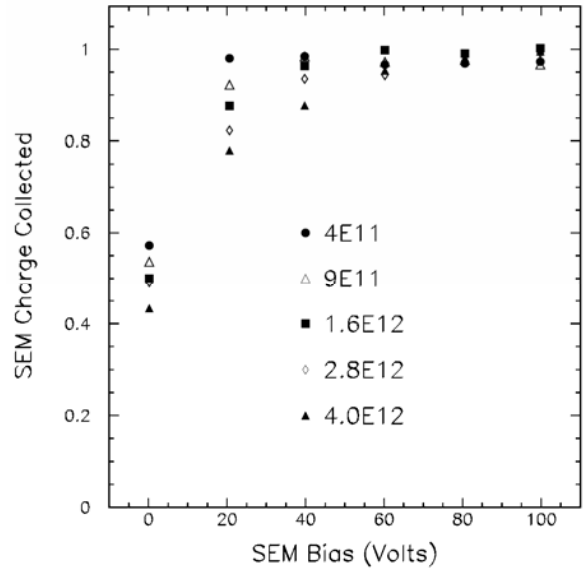

**FIGURE 5.** Fraction of SEM charge collected *vs* the bias voltage. The vertical scale is normalized to 1.0 for the data point at 100V and $4\times10^{11}$/spill. The study is repeated for 5 beam intensities.

The Ti foil SEM experiences 12μm elongation. Because of the 4 mm extension in the accordion springs, we note that this elongation results in <1% tension loss. Of note as well is the fact that a W-wire SEM, when strung on a frame at the maximum stress (the yield stress), stretches the wires by 160μm; thus, beam heating will drop the tension of such a wire SEM by nearly ⅓. Such loss motivates the accordion springs for the foil SEM or individual springs on wire SEM's as has been used by previous workers [16].

## BEAM TEST RESULTS

A prototype foil SEM detector was tested in the 8 GeV transfer line from the FNAL Booster to the MiniBoone experiment. The 8 GeV beam has $\sigma_{beam}\approx 3$ mm at the location of the SEM, and the spill is 1.56 μsec in duration. The beam intensity was varied from $4\times10^{11}$/spill to $4\times10^{12}$/spill. A BPM was available within 3m of the SEM and the beam intensity was measured at Booster extraction with a BCT.

The first test was to measure the signal yield off the SEM as a function of voltage on the bias foils. The results are shown in Figure 5 for several spill intensities. The vertical axis in Figure 5 is the fractional charge collected, normalized to be 1.0 for the data point at 100Volts and $4\times10^{11}$/spill intensity. The data suggest a need for larger applied voltage as the beam intensity increases.

The second purpose of this test was to study the SEM resolution for determining the beam centroid. This was accomplished by combining two studies, one in which beam was at a stable position, and one in which it wandered over the SEM.

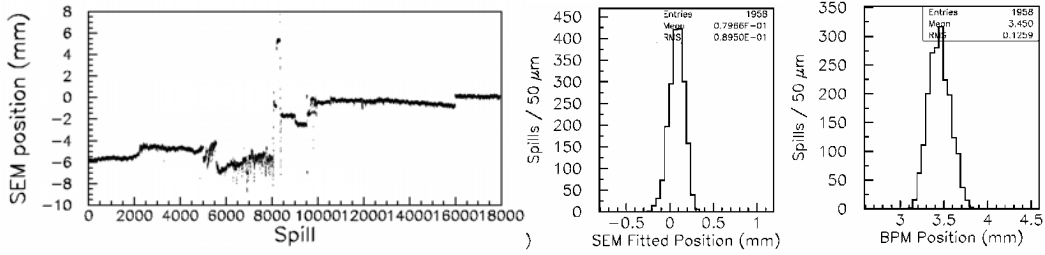

**FIGURE 6.** (left) Beam centroid position as measured by the SEM *vs* spill number for a several hour period. (middle) Projection of the left plot over the spill range 16000-18000. (right) Histogram of the BPM centroid measurement for the same range of spills as the middle plot for the SEM.

The results of the first resolution study are shown in Figure 6. The beam appears to be stable for spills 16000-18000. A histogram of the SEM and BPM centroids over this spill range are shown in the middle and right graphs of Figure 6, showing RMS values of 90 μm and 125μm, respectively. These magnitudes are a combination of the detector resolutions as well as the extent to which the beam wanders. We have:

$$\sigma_{SEM}^2 + \sigma_{wander}^2 = (90\mu m)^2 \quad \text{and} \quad \left(\sigma_{BPM}^2 + \sigma_{wander}^2\right)\alpha^2 = (125\mu m)^2 \qquad (1)$$

The factor α=1.13 is a scale factor between the SEM and BPM, and reflects that the calibration of the two detectors may be imperfect; *ie*: one millimeter in the SEM may be more than a millimeter as read out by the BPM. The value of α is obtained below. To factor out the effect of beam wandering, we take the difference of these results:

$$\sigma_{BPM}^2 - \sigma_{SEM}^2 = (125\mu m/\alpha)^2 - (90\mu m)^2 \qquad (2)$$

In the second resolution study, we correlated the fitted SEM centroid with the BPM. The upper two plots of Figure 7 show the BPM and SEM position for 6499 spills histogramed over a several hour period. During this period, the beam wandered by ~2 mm, as is seen in both detectors. These plots likewise suggest $\sigma_{SEM} < \sigma_{BPM}$.

The lower left plot of Figure 7 shows the correlation of the BPM and SEM positions and a line fit to the data. The lower right plot of Figure 7 shows the residuals of these data with respect to the fitted line. The RMS of the residuals, which has contributions from both the BPM and SEM resolutions, is:

$$\alpha^2 \sigma_{BPM}^2 + \sigma_{SEM}^2 = (120\mu m)^2 \qquad (3)$$

The scale factor α here may be obtained from the slope of the scatter plot of SEM vs BPM centroid in Figure 7; we find α=1.13 Combining Equations (2) and (3),

$$\sigma_{SEM} = 62\mu m \quad \text{and} \quad \sigma_{BPM} = 92\mu m \qquad (4)$$

These results, achieved with a 1 mm pitch SEM in a 3 mm beam, suggest that the SEM will be adequate to achieve 100 μm resolution in the 1-2 mm NuMI beam.

## ACKNOWLEDGMENTS

It is a pleasure to thank M. Raizen and K. Shih, of the University of Texas, B. Baller, S. Childress, D. Harris, and C. Kendziora of FNAL, and especially G. Ferioli and J. Camas of CERN for helpful conversations. We thank R. Zwaska, T. Kobilarcik, C. Moore, and G. Tassotto for assistance with the beam test. This work was supported by the U.S. Department of Energy and the Fondren Family Foundation.

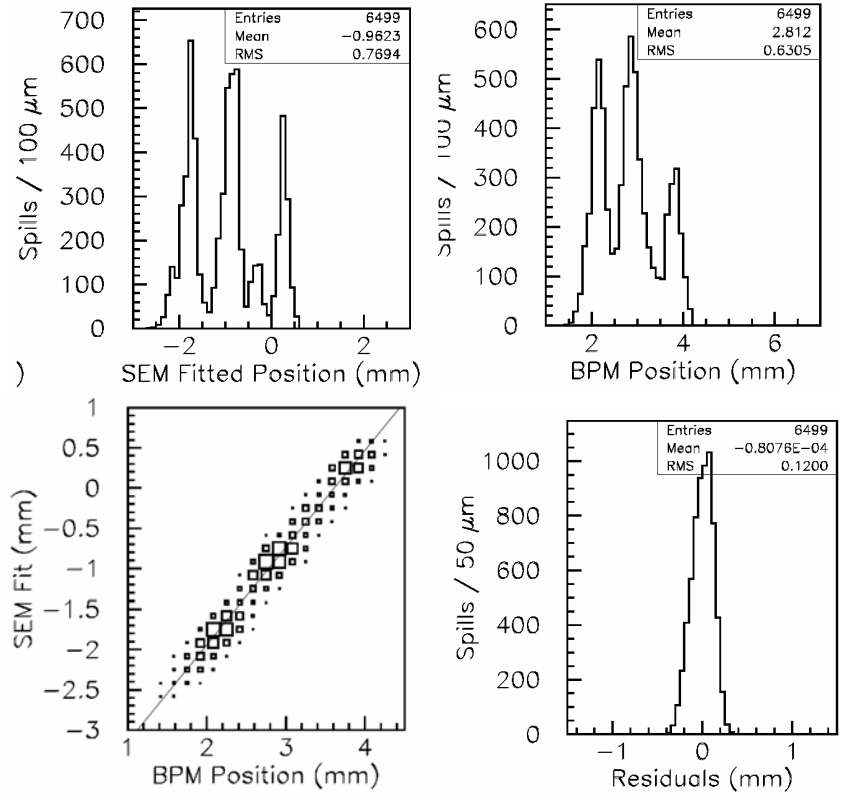

**FIGURE 7.** (upper left) SEM centroid position for a few hour time period in which beam center move by nearly 3 mm. (upper right) Same info as reported by the nearby BPM; (lower left) Scatter plot of SEM and BPM centroid positions, along with a best fit line; (lower right) Residuals of the plot at left.